\documentstyle[preprint,aps,epsf]{revtex}

\tightenlines

\begin{document}
\draft
\title { Disappearance of Elliptic Flow:
A New Probe for the Nuclear Equation of State  }
\author{
                P. Danielewicz$^2$, Roy~A.~Lacey$^1$, P.-B.~Gossiaux$^{2,3}$,
C.~Pinkenburg$^1$,               P.~Chung$^1$,  J.~M.~Alexander$^1$, and
R.~L.~McGrath$^1$\\
                         }
\vskip 1cm
\address{$^1$Departments of Chemistry and Physics, State University of
New York at Stony Brook,  \\
Stony Brook  NY 11794-3400, USA \\
$^2$National Superconducting Cyclotron Laboratory and\\
the Department of Physics and Astronomy, Michigan State University, \\
East Lansing MI 48824-1321, USA \\
$^3$SUBATECH, Ecole des Mines, F-44070 Nantes, France \\
}
\date{\today}
\maketitle
%
%
\begin{abstract}
        Using a relativistic hadron transport model, we investigate the utility
of the elliptic flow excitation function as a probe for the stiffness of
nuclear matter and for the onset of a~possible quark-gluon-plasma (QGP)
phase-transition at AGS energies  $1 \lesssim E_{Beam} \lesssim 11$~AGeV.
The~excitation function shows a~strong dependence on the nuclear equation of
state, and exhibits
characteristic signatures which could signal the onset of a~phase transition to
the QGP.
\end{abstract}


  The delimitation of the parameters of the nuclear equation of state~(EOS) has
been, and continues to be, a~major driving force for studies of nuclear matter
at high energy densities~\cite{ber88,shu80}. A~prevailing notion is that such
studies will not only provide crucial information about  the~EOS, but will
afford unique information about novel non-perturbative aspects of quantum
chromodynamics (QCD)~\cite{shu80,gro81}.  The~phase transitions for
deconfinement and chiral symmetry restoration  leading to quark-gluon
plasma~(QGP) formation  constitute aspects which are of great current interest.

        A~key step toward the delineation of the EOS and the QGP phase
transitions is the determination of experimental signatures which can serve as
clean and sensitive probes.  Many of the experimental signatures proposed over
the last several years~\cite{sie79,mat86,dov95} have had only limited success,
primarily because they have turned out to be sensitive not only  to the EOS or
the phase transition, but also to other poorly known physical effects.

Recent lattice quantum chromodynamics~(QCD) calculations (carried out at zero
baryon density) point to a~softening of the~EOS near and above the energy
region of the QGP phase transition, with the pressure rising with temperature
more slowly than the energy density.  This softening starts  even for
quark-antiquark densities quite comparable to those in the ground-state of
nuclear matter~\cite{lae96}. In~the picture where changes in the properties of
hadronic matter are associated with hadrons pushing out the nonperturbative
vacuum, one expects a~similar softening of the~EOS
even at relatively low temperatures when the baryon density increases
significantly beyond its normal value~$\rho_0$. Collisions of nuclei at AGS
energies ($1 \lesssim E_{Beam} \lesssim 11$~AGeV) are expected to produce
equilibrated matter at densities up to $\rho \sim 8 \rho_0$.  Thus, it is of
paramount importance to establish signatures  for a~softening of the~EOS for
such matter densities.

\begin{figure}
\centerline{\epsfysize=7.3in \epsffile{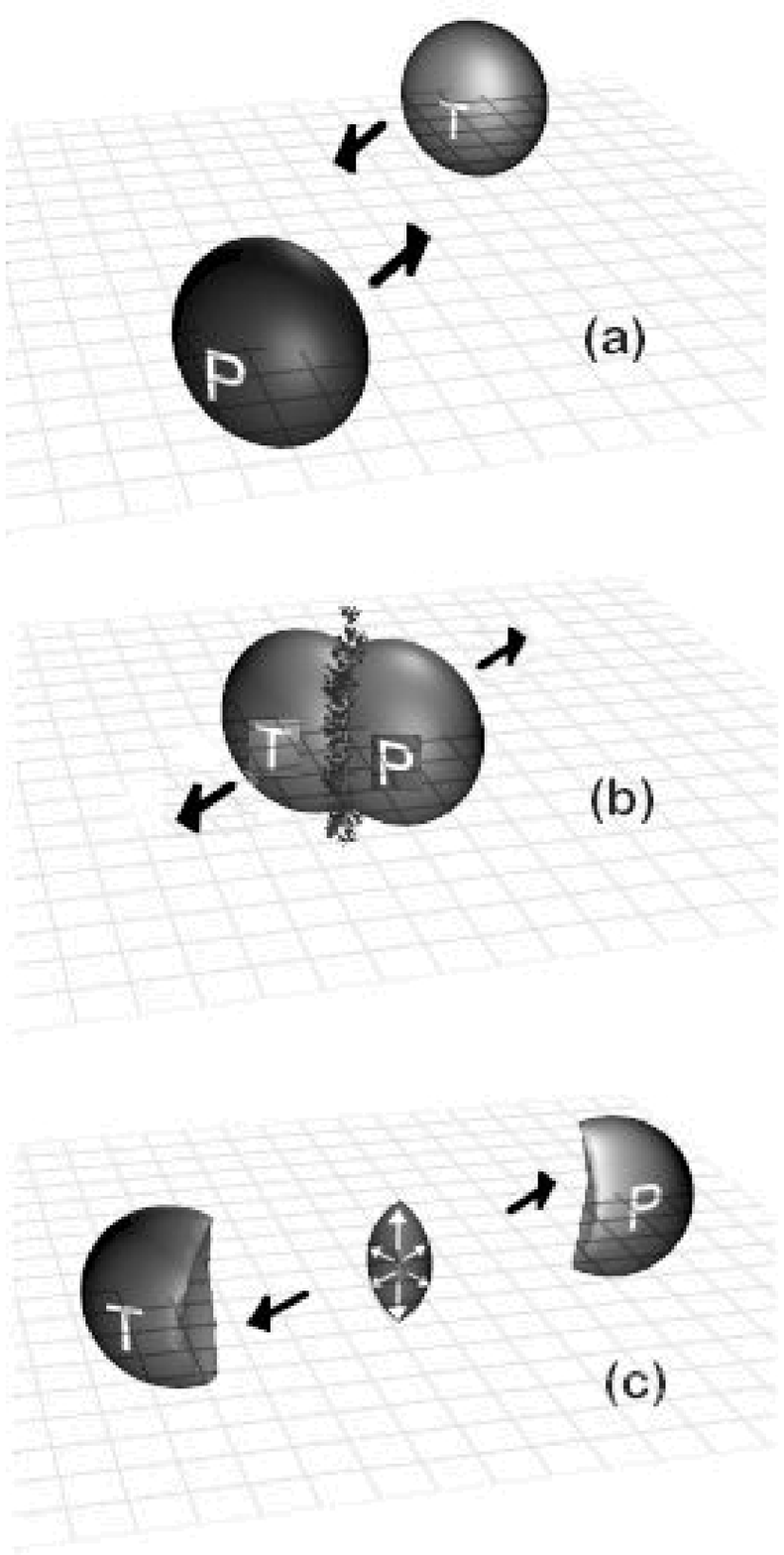}}
\vspace*{.4in}
\caption{
                Schematic illustration of the collision of
two Au nuclei at relativistic energies.   Time shots are shown for an~instant
before the
collision~(a), early in the collision~(b), and late in the collision~(c).}
\label{fig1}
\end{figure}

        Pressure gradients provide the driving force for collective flow and
the associated
anisotropy of the transverse flow tensor.  Consequently,  flow measurements
can provide useful probes for the pressure  developed in nuclear collisions,
and hence,  may provide useful insights into the EOS as well as QGP formation.
The~directed sideward flow signal at relativistic energies is compromised
by the fact that the flow angle esentially merges with the collision axis. In
addition, this signal is known to exhibit  strong dependencies on transport
properties~\cite{dan95,dan88}  which become increasingly uncertain with
increasing beam
energy.   These~dependencies are weaker for  the transverse flow tensor
(elliptic flow) which has already been shown to be quite sensitive to the
pressure at maximum compression~\cite{oll92,sor97}. At~AGS energies of  $\sim 1
- 11$ AGeV, the~elliptic flow results from a~strong competition between
squeeze-out and
in-plane flow.  In~the early stages of the collision as shown in
Fig.~\ref{fig1}(b), the spectator nucleons block the path of participant
hadrons emitted toward the reaction plane;  therefore the nuclear matter is
initially squeezed out preferentially orthogonal to the reaction plane. This
squeeze-out of nuclear matter leads to negative elliptic flow.  In the later
stages of the reaction as shown in Fig.~\ref{fig1}(c), the~geometry of the
participant region (i.e.~a~larger
surface area exposed in the direction of the reaction plane)  favors in-plane
preferential emission and hence positive elliptic flow.  The~magnitude and the
sign of
the elliptic flow depend on two factors: (i)~the pressure built up in the
compression stage compared to the energy density, and (ii)~the passage time for
removal of the shadowning due to the projectile and target spectators.
Specifically, the~characteristic time for the development of expansion
perpendicular to the reaction plane is $\sim R/c_s$, where the speed
of sound is $c_s =\sqrt{\partial
p / \partial e}$, $R$ is the nuclear radius, $p$ is the pressure and $e$ is the
energy density. The passage time is $\sim 2R /(\gamma_0 \, v_0)$, where $v_0$
is the c.m. spectator velocity.  The~squeeze-out  contribution should then
reflect  the ratio~\cite{dan95}
\begin{equation}
{c_s \over \gamma_0
\, v_0 } \, .
\label{ratio}
\end{equation}
Elliptic flow measurements can therefore give a probe for  $c_s $, and their
excitation functions can give a good probe for the EOS.

                Theoretical efforts utilizing only hydrodynamical concepts have
provided a wealth of qualitative insights on the nature of collective flow.
However, these approaches often lead to serious overestimates of the
quantitative strength of various dynamical effects~\cite{dan88,oll92,ris96}.
In this Letter, we investigate -- by way of a~relativistic transport model
--the practical viability of elliptic flow as a probe for the EOS and the QGP
in Au + Au collisions at AGS energies ($ \sim$~1 -11 AGeV).  We show that:
(i)~a~distinct
transition from negative to positive elliptic flow occurs at a~beam energy,
E$_{tr}$, which is strongly dependent on the parameters of the EOS and
(ii)~that a~phase transition to the QGP would provide a  characteristic and
detectable signature in the elliptic flow excitation function.

                Calculations have been performed with a~preliminary version of
our~new
relativistic transport model~\cite{gos98}.  An~ultimate goal for this new
effort is to have a model which is not only applicable at finite baryon
densities and in nonequilibrium situations, but can reproduce the thermodynamic
properties of nuclear matter obtained in lattice QCD calculations.
The~phenomenological relativistic Landau theory of quasiparticles~\cite{bay76}
serves as a~basis for the model. For the purpose of this Letter, which is
concerned only with energies of $\sim 1 - 11$~AGeV, we choose nucleons, pions,
and $\Delta$ and $N^*$ resonances for the hadronic degrees of freedom. Particle
energies are
evaluated by specifying the energy density (in any frame) as a~functional of
the particle phase-space distributions.  The~fields are chosen here solely
to~act  on baryons as these
particles offer the best chance to observe any emission anisotropy  that is
of a dynamic origin.  Since the vector and scalar fields can be momentum
dependent with no exclusive dependence on the vector or scalar densities, there
is no particular benefit  for a~separate consideration of such
fields (unless, of course, the spin dynamics were considered).

\begin{figure}
\centerline{\epsfxsize=6.0in \epsffile{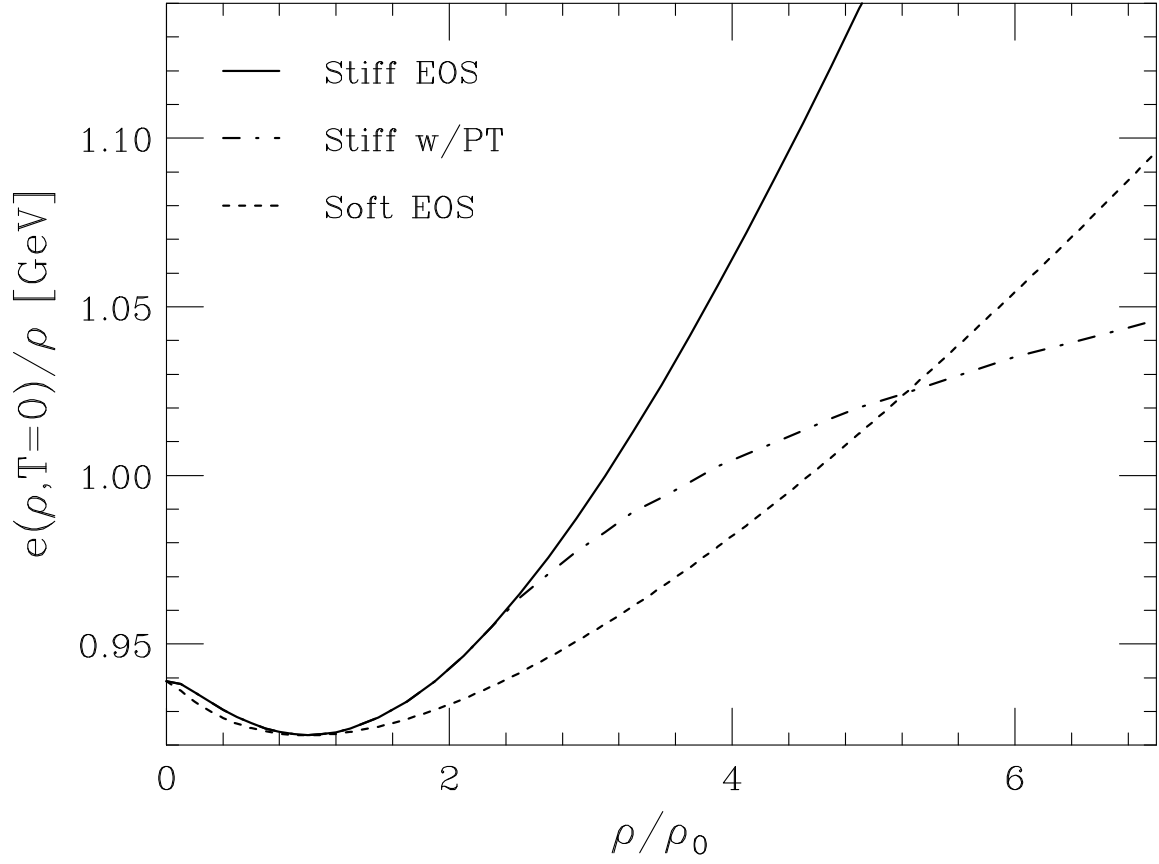}}
\vspace*{.4in}
\caption{
                    Energy per baryon vs.\  baryon density, at $T = 0$. Curves
are shown for a~stiff EOS (solid curve), a~soft EOS
(dashed),
and an~EOS with a~second-order phase transition (dashed-dot).
}

\label{fig2}
\end{figure}

        For convenience, we choose a~parametrization for the fields that could
be
easily identified as either vector or scalar\cite{pan93}. Thus, in the case of
the fields without momentum dependence in their nonrelativistic reduction, we
use a baryon energy density of the form
\begin{equation}
e = \sum_X {g_X \over (2 \pi)^3 } \int d{\bf p} \, f_X ({\bf
p}) \, \sqrt{p^2 + m_X^2 (\rho_s)}
+ \int_0^{\rho_s} d{\rho_s}' \, U(\rho_s') -
\rho_s \, U(\rho_s),
\label{e=}
\end{equation}
where the summation is over all baryons, $m_X (\rho_s) = m_X +
U(\rho_s)$, and $\rho_s~=~\sum_X {g_X \over (2 \pi)^3} \int
d{\bf p} \, {m_X (\rho_s) \over \sqrt{p^2 + m_X^2 (\rho_s)}} \, f_X ({\bf p})$.
 This gives baryon single-particle energies
\begin{equation}
\epsilon_X (p, \rho_s) = \sqrt{p^2 + m_X^2(\rho_s)}.
\end{equation}
We take
\begin{equation}
U({ { \xi}}) = {-a \, {{\xi}} + b \, {{\xi }}^\nu \over 1 +
({{ \xi }}/2.5)^{\nu - 1}} \, ,
\label{U=}
\end{equation}
where ${{\xi }} = \rho_s / \rho_0$, $\rho_0 = 0.16$~fm$^{-3}$, and $a$, $b$,
and $\nu$ are adjusted to the ground-state nuclear matter properties:
$a=187.24$~MeV, $b=102.62$~MeV, and $\nu = 1.5354$ for a soft
equation of state
($K=210$~MeV), and $a=115.08$~MeV, $b=48.25$~MeV, and $\nu=2.5427$ for a~stiff
equation of state~($K=380$~MeV). It is important to point out here that the
denominator in~(\ref{U=}) prevents supraluminous behavior at high densities.
The~resulting energy per baryon at $T=0$, for the two~EOS, is plotted as
a~function of baryon density in Fig.~\ref{fig2}.  The energy
per baryon shows a~steeper increase with density for the solid curve (the stiff EOS) than for the
dashed line (the soft EOS).  This  leads to higher pressures and sound
velocities for the stiff EOS.

        In the case of mean-fields dependent on momentum in their
nonrelativistic reduction, we parametrize the energy density in
the local frame where ${\bf J} = \sum_X {g_X \over (2 \pi)^3}
\, \int d{\bf p} \, f_X ({\bf p}) \, {\partial \epsilon_X \over \partial {\bf
p}} = 0$, with
\begin{equation}
e = \sum_X {g_X \over
(2 \pi)^3 } \int d{\bf p} \, f_X ({\bf
p}) \left( m_X + \int_0^p dp' \, v_X^*(p', \rho) \right) +
\int_0^\rho d\rho ' \, U(\rho '),
\end{equation}
and  $U$ is of the form expressed by Eq.~(\ref{U=}) with
${{ \xi }} = \rho / \rho_0$ and $\rho = \sum_X  \,  {g_X \over
(2 \pi)^3} \, \int d{\bf p} \, f_X ({\bf p}) $.  Here the
local particle velocity $v_X^*$ is of the form
\begin{equation}
v_X^*(p,{{ \xi }}) = {p \over \sqrt{p^2 + m_X^2
\left/ \left( 1 + c \, {m_N \over m_X}  \, {{{ \xi }} \over (1
+ \lambda \, p^2/m_X^2)^2} \right)^2 \right. } } \, .
\end{equation}
This yields single-particle energies
\begin{equation}
\epsilon_X(p, \rho) = m_X + \int_0^p dp' \, v_X^* +
 \rho
\left\langle \int_0^p dp' \, {\partial v \over \partial \rho}
\right\rangle + U(\rho) \, ,
\label{eps=}
\end{equation}
which in a nonrelativistic reduction is similar to the energies in the
nonrelativistic transport models proposed by Bertsch, Das Gupta {\em et
al.}~\cite{ber88,cse92}.  (For completeness, it should be noted that  for
matter
without local reflection symmetry in momentum space, there is
a~correction term
in Eq.~(\ref{eps=})  from ${\bf J} = 0$ , that we ignore here; it is
interesting
that this term is never mentioned in connection with nonrelativistic
calculations.)
Adjustment of the parameters to the properties of nuclear matter and to the
nucleon-nucleus potential gives: $a = 185.47$~MeV, $b = 36.29$~MeV, $\nu =
1.5391$, $c =0.83889$, and $\lambda = 1.0890$ for a~soft equation of state
($K=210$~MeV), and $a = 123.62$~MeV, $b = 14.65$~MeV, $\nu = 2.8906$,
$c=0.83578$, and $\lambda = 1.0739$ for a~stiff equation ($K=380$~MeV).
The effective mass at normal density and Fermi momentum is
$m_N^* = {p_F / (\partial \epsilon_N / \partial p)} = 0.65 \,m_N$ for both
sets.

        The equations of motion in the mean field have been integrated
utilizing a~relativistic Hamiltonian for the computational lattice. When
addressing
production processes in collisions of heavy nuclei, it is important that the
rates obey detailed balance for processes which take place in the
vicinity of equilibrium~\cite{dan91}. This is a relatively straightforward
condition to satisfy for two-body collisions and for resonance formation and
decay. However, it is difficult for processes with three or more particles in
the initial or final states.  Based on the average multiplicities of produced
particles and on the typical conditions of equilibrium, we adopt a~compromise
in our transport model by dividing the elementary hadron
collision-processes
into low- and high-energy ($\sqrt{s} - m_1 - m_2 > 1.7$~GeV) processes.
The~elementary low-energy processes, with at most two particles in any state,
have a strictly enforced detailed balance, in contrast to the high-energy
processes for which the inverse processes are less likely.  The high-energy
production processes have been parametrized using available experimental data
on net cross sections, pion multiplicities, hadron rapidity and
transverse-momentum distributions (e.g.~\cite{gra78,bud80}); these processes
are simulated using the concept of transverse-momentum phase-space with
a~leading particle effect. The procedure is  similar to that of
Ref.~\cite{pan93a}. When possible, the diffractive production is favored over
the central.  The~low-energy processes are treated in a relatively standard
manner~\cite{dan91}. The~model, so-constructed, successfully reproduces pion
and proton rapidity distributions as well as transverse-momentum distributions
within the beam-energy range of interest.

        A~necessary prerequisite for using elliptic flow as
a~probe for the EOS
and the QGP is that a~chosen measure of this flow can
accurately determine both
its magnitude and sign. We use the Fourier coefficient $\langle \cos{2 \phi}
\rangle$, to measure the elliptic azimuthal asymmetry of the particle
distributions at midrapidity ($|y_{cm}| < 0.3\,y_0$). In the expression for
the coefficient, $\phi$~represents the azimuthal angle of an~emitted baryon
relative to the reaction plane.  In~the absence of elliptic
flow, the~particle
distribution is isotropic and this coefficient vanishes.
For~in-plane elliptic
flow (i.e.\ positve) the Fourier coefficient $\langle \cos{2
\phi} \rangle > 0$
and for  out-of-plane elliptic flow (i.e.\ negative ) $\langle
\cos{2 \phi}
\rangle < 0$.  For small values of the ratio ${c_s / \gamma_0 \, v_0 }$,
the~squeeze-out (or negative) contribution should decrease approximately
linearly with  this ratio.

\begin{figure}
\centerline{\epsfxsize=6.0in \epsffile{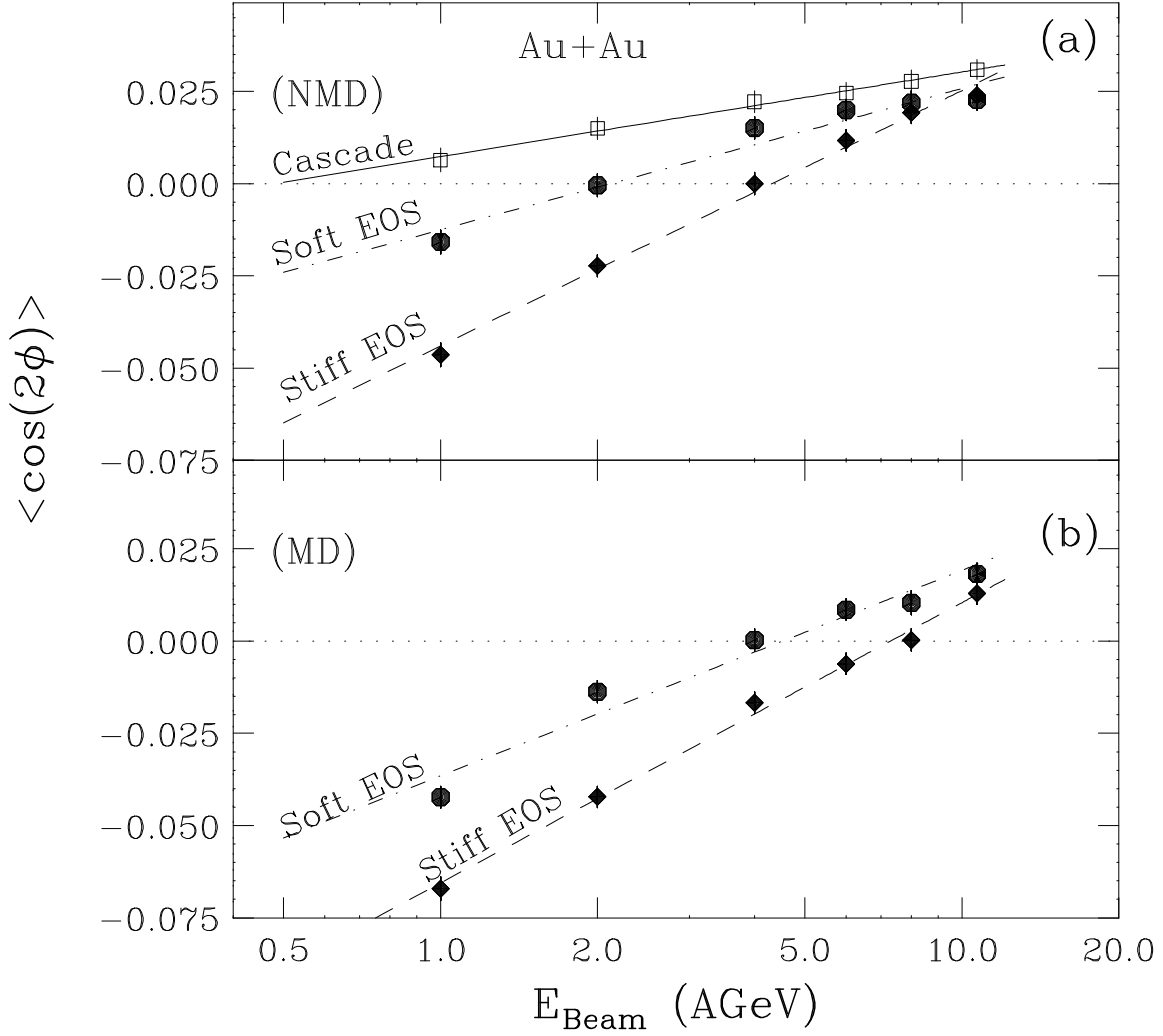}}
\vspace*{.4in}
\caption{
                        Calculated elliptic flow excitation
functions for Au + Au reactions.
Panels~(a) and~(b) show, respectively, the~functions obtained
without~(NMD) and with~(MD) the momentum dependent forces.
The~filled circles, filled diamonds, and open squares indicate,
respectively, results obtained using a~soft EOS, a~stiff EOS,
and by neglecting the mean field.  The~straight lines show logarithmic
fits.
}
\label{fig3}
\end{figure}

        An important characteristic of the elliptic flow is the change from
negative values at low beam energies ($ \sim $0.2~AGeV) to positive values at
high beam energies ($ \sim $20~AGeV).
This behavior is clearly exhibited in Fig.~\ref{fig3}(a) where we show elliptic
flow excitation functions  for calculations which assume a stiff  EOS (filled
diamonds), and  a~soft EOS (filled circles).  For comparison we also show
results  from calculations performed without a mean field. All of the results
shown in Fig.~\ref{fig3}(a) are calculated for  impact parameters of   $b =
4-6$~fm. Figure~\ref{fig3}(a)  shows an essentially logarithmic
dependence of the
elliptic flow on beam energy over the range $0.5
\lesssim E_{Beam} \lesssim 10$~AGeV. This trend is the expected  qualitative
result  and points to the fact that the~squeeze-out contribution to the
elliptic flow decreases with increasing beam energy.
Figure~\ref{fig3}(a) also
shows  that the slope of the excitation function, as well as the beam energy
$E_{tr}$,  for  transition from negative to positive elliptic flow,
depends strongly  on the stiffness of the~EOS.  We believe that these results
not only  provide a new and significant probe for the EOS, but raise the
expectation that any change in the stiffness of the~EOS with energy, will be
manifested as a change in the slope of the excitation function.

        Experience from the low-energy domain ($E_{Beam} \lesssim 1$~AGeV) has
shown that the~momentum dependence of the mean field is an important factor for
the
determination of the stiffness of the EOS from flow measurements.
Nucleon-nucleus scattering experiments and  nuclear-matter calculations clearly
indicate the presence of this momentum dependence; here, it plays a~role in
generating flow before the hadronic matter equilibrates.  A priori,
there could  be two separate effects due to the momentum
dependence: (i)~it may
pass for an~enhanced stiffness of the ~EOS, and/or (ii)~it may
lead to a loss of
sensitivity to the stiffness of the EOS .  The~second effect could eliminate
the possibility of observing a~change in the stiffness with increasing  beam
energy.

        In Fig.~\ref{fig3}(b), we show elliptic flow excitation functions
obtained
from calculations that include momentum-dependent fields.  The general trends
of these excitation functions are similar to those shown in Fig.~\ref{fig3}(a).
However, one can clearly see that the net effect of the momentum dependence is
to enhance the squeeze-out.  Of greater significance is the fact that
the~sensitivity of the elliptic flow to the stiffness of the EOS remains
practically unchanged when this momentum dependence is included. The~difference
in the transition energy~$E_{tr}$ is $\sim$2~AGeV, between the soft and stiff
EOS whether or not one includes momentum dependent forces (see
Figs.~\ref{fig3}).  A cursory examination of
Fig.~\ref{fig3} also shows that measurements of elliptic flow should clearly
discriminate between models with a~realistic EOS and the cascade models for
which $E_{tr} \lesssim 0.6$~AGeV.  The latter result corroborates an~earlier
finding on the second-order flow~\cite{tsa96}.

\begin{figure}
\centerline{\epsfxsize=6.0in \epsffile{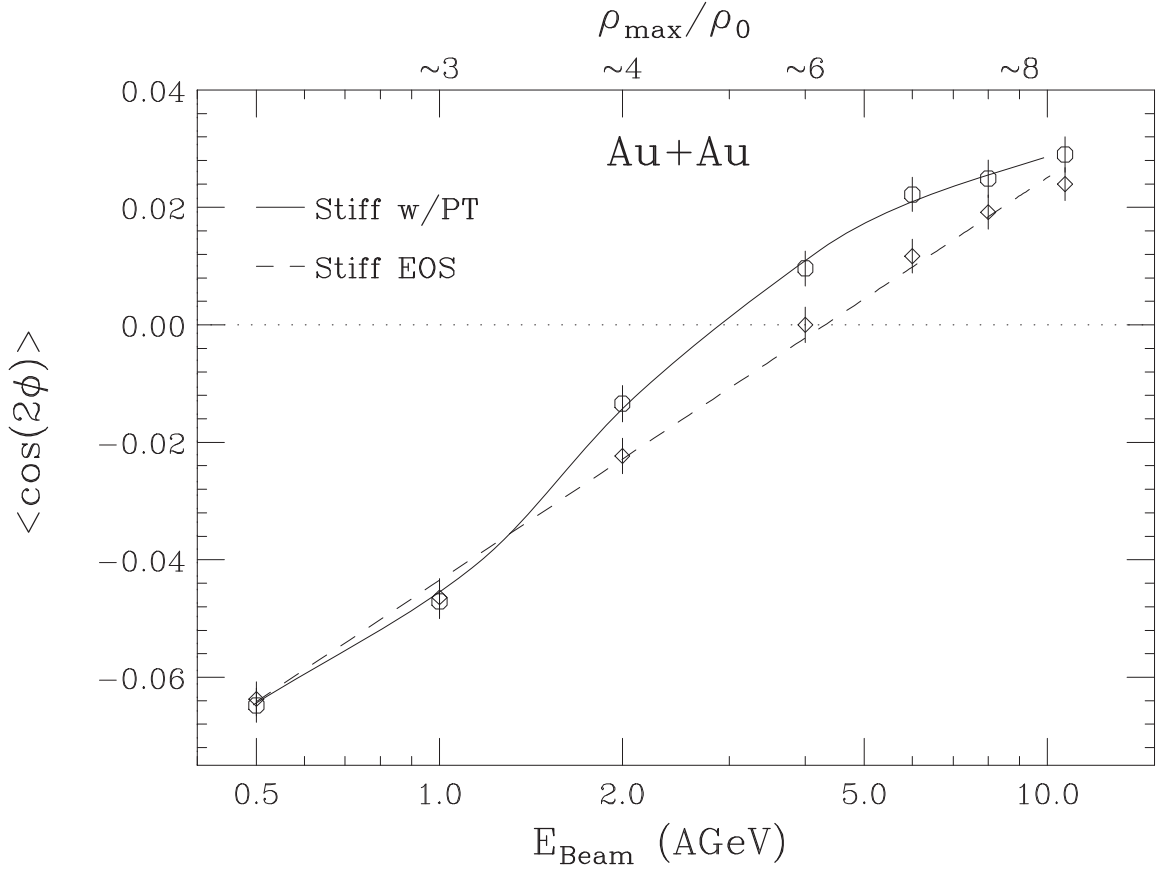}}
\vspace*{.4in}
\caption{
                      Calculated elliptic flow excitation functions for Au +
Au. The~open diamonds represent results obtained with a~stiff EOS. The~open
circles represent results obtained with a~stiff EOS  and with a~second-order
phase transition (see text).  The~solid and dashed lines are drawn to
guide the eye.  Numbers at the top of the figure indicate rough magnitude
of local baryon densities reached in reactions at different beam energies.
}
\label{fig4}
\end{figure}

        In order to search for an elliptic-flow signature that can signal the
onset of
a phase transition to the QGP, we have carried out calculations assuming an~EOS
with a~weak second-order phase transition, obtained by setting $U
=\mbox{const}$ in (\ref{e=})-(\ref{U=}), for~$\xi \ge 2.3$.
The~energy per baryon for this EOS is shown by the dash-dotted curve
in~Fig.~\ref{fig2}. A~first-order phase transition with a~stronger drop in the
energy per baryon (in the same general density range) is obtained in the
extrapolation of the lattice QCD results~\cite{gos98} .
The~elliptic-flow excitation functions calculated using  a stiff EOS with a
phase transition
(open circles) and a stiff EOS without the phase-transition (diamonds) are
compared in Fig.~\ref{fig4}.  Both functions have been obtained with no
consideration of momentum dependence in the mean field.  For low beam energies
($\lesssim \, 1$~AGeV), the~elliptic flow excitation functions are essentially
identical because the two EOS are either identical or not very different at the
densities and temperatures that are reached.  For $2 \lesssim
E_{Beam} \lesssim 9$~AGeV the~excitation function shows  larger in-plane
elliptic flow from the calculation which includes the phase transition ,
indicating that a~softening of the EOS has occured for this beam energy range.
This deviation is in direct contrast to the esentially logarithmic  beam energy
dependence obtained (for the same energy range) from the calculations which
assume a~stiff EOS without the phase transition.  This difference in the
predicted excitation functions could very well serve as an~important hint as to
whether or not the conditions necessary for the phase transition to the QGP are
created  in this energy range.

        In summary, we have used a~relativistic hadron transport model to
investigate the viability of elliptic flow as a~probe for the compressibility
of nuclear matter and the onset of a possible QGP phase transition in the
energy range $1 - 11$~AGeV . The~flow excitation functions
obtained from our model calculations indicate that a~distinct transition from
out-of-plane elliptic flow to in-plane elliptic flow occurs at a~beam energy,
E$_{tr}$, which is strongly dependent on the incompressibility constant for
nuclear matter. Additional calculations, which take account of a~possible phase
transition, show characteristic trends in the excitation functions for elliptic
flow, which could signal the onset of a~QGP phase transition. These new results
strongly suggest that an experimental analysis  of elliptic flow from reaction
data  at AGS energies can provide invaluable constraints for the EOS and
possibly for formation of the QGP.

\acknowledgements
        This work was supported in part by the U.S.\ Department of
Energy under Contract No.\ DE-FG02-87ER40331.A008 and by the
National Science Foundation under Grant No.\ PHY-96-05207.

\newpage

\end{document}